\documentclass[12pt]{article}
\usepackage{amssymb}
\usepackage{epsf}
\usepackage{lscape}
\usepackage[cp866]{inputenc}
\usepackage[T2A]{fontenc}
\usepackage[english]{babel}
\begin{document}
%\draft
\title{\Large\bf {Determination of the $ ^3{\rm {He}}+\alpha \to^7\rm {Be}$ asymptotic
normalization coefficients (nuclear vertex constants) and their
application for extrapolation of the $^3{\rm
{He}}(\alpha,\gamma)^7{\rm {Be}}$ astrophysical $S$-factors
 to  the solar energy region }}
%(\today)
\author{S.B. Igamov, K.I. Tursunmakhatov  and R. Yarmukhamedov \thanks{Corresponding author, E-mail: rakhim@inp.uz}}
\maketitle {\it {Institute of Nuclear Physics, Uzbekistan Academy of
Sciences,100214 Tashkent, Uzbekistan}}

 \begin{abstract}
 A new analysis of the precise experimental astrophysical $S$-factors
for the direct capture $^3He(\alpha,\gamma)^7{\rm {Be}}$ reaction
[B.S. Nara Singh et al., Phys.Rev.Lett. {\bf 93} (2004) 262503; D.
Bemmerer et al., Phys.Rev.Lett. {\bf 97} (2006) 122502;
F.Confortola et al., Phys.Rev. {\bf C 75} (2007) 065803  and
T.A.D.Brown et al., Phys.Rev. {\bf C 76} (2007) 055801] populating
to the ground and first excited states of $^7{\rm Be}$ is carried
out based on the modified two - body potential approach in which
the direct astrophysical $S$-factor, $S_{34}(E)$, is expressed in
terms of the asymptotic normalization constants for $^3{\rm
{He}}+\alpha\to ^7{\rm {Be}}$ and two additional conditions are
involved to verify the peripheral character of the reaction under
consideration. The Woods--Saxon potential form is used for the
bound ($\alpha+^3{\rm {He}}$)- state and the $^3{\rm {He}}\alpha$-
scattering wave functions.   New estimates are obtained for the
$^{\glqq}$indirectly measured\grqq\, values of the asymptotic
normalization constants (the nuclear vertex constants) for $^3{\rm
{He}}+\alpha\to^7{\rm {Be}}(g.s.)$ and $^3{\rm
{He}}+\alpha\to^7{\rm {Be}}(0.429 MeV)$ as well as the
astrophysical $S$-factors $S_{34}(E)$ at E$\le$ 90 keV, including
$E$=0. The  values of asymptotic normalization constants have been
used for getting information about the $\alpha$-particle
spectroscopic factors for the mirror ($^7Li^7{\rm {Be}}$)-pair.
\end{abstract}


\section{Introduction}
\hspace{0.7cm}  The  $^3{\rm {He}}(\alpha,\gamma)^7{\rm {Be}}$
reaction is one of the  critical links in the $^7{\rm {Be}}$ and
$^8B$ branches of the $pp$--chain of solar hydrogen burning [1--3].
The total capture rate determined by processes of this chain is
sensitive to the cross section $\sigma_{34}(E)$ (or the
astrophysical $S$-factor $S_{34}(E)$ ) for the $^3{\rm
{He}}(\alpha,\gamma)^7{\rm {Be}}$ reaction and predicted neutrino
rate varies as $[S_{34}(0)]^{0.8}$  \cite{Bah82,Bah92}.

Despite the impressive improvements in our understanding of the
$^3{\rm {He}}(\alpha,\gamma)^7{\rm {Be}}$ reaction  made in the past
decades (see Refs [4--10] and references therein), however, some
ambiguities connected with both the extrapolation of the measured
cross sections for the aforesaid reaction to the solar energy region
and the theoretical predictions for $\sigma_{34}(E)$ (or
$S_{34}(E)$) still exist and they may  influence the predictions of
the standard solar model \cite{Bah82,Bah92} .

Experimentally, there are two types of data for the $^3{\rm
{He}}(\alpha,\gamma)^7{\rm {Be}}$ reaction at extremely low
energies: i) six measurements based on detecting of $\gamma$-rays
capture  (see \cite{Adel98} and references therein) from which the
astrophysical $S$-factor $S_{34}(0)$ extracted by the authors of
those works changes within the range 0.47$\le S_{34}(0)\le$0.58
${\rm {keV\,\,\, b}}$, which  yield a weighted mean of
$S_{34}(0)$=0.507$\pm$ 0.016 ${\rm  {keV\,\,\, b}}$ \cite{Adel98},
and  ii) five measurements based on detecting of $^7{\rm {Be}}$ (
see \cite{Adel98} references therein as well as [6-10]) from which
$S_{34}(0)$ extracted by the authors of these works changes within
the range 0.53$\le S_{34}\le$0.63 ${\rm  {keV\,\,\,b}}$ , which
yield weighted means of $S_{34}(0)$=0.572$\pm$ 0.026 ${\rm
{keV\,\,\,b}}$ \cite{Adel98}, $S_{34}(0)$=0.53 ${\rm  {keV\,\,\,b}}$
\cite{Nara04}, $S_{34}(0)$=0.547$\pm$ 0.017 ${\rm  {keV\,\,\,b}}$
\cite{Bem06,Gy07}, and $S_{34}(0)$=0.560$\pm$ 0.017 ${\rm
{keV\,\,\,b}}$ \cite{Con07} and $S_{34}(0)=$0.595$\pm$ 0.018 and
0.596$\pm$ 0.021 ${\rm  {keV\,\,\,b}}$ \cite{Brown07}. All of these
measured data have a similar energy dependence for the astrophysical
$S$-factors $S_{34}(E)$ but the extrapolation of  each of the
measured data from the observed energy ranges to low experimentally
inaccessible energy regions, including $E$=0, gives a value of
$S_{34}(0)$ with an uncertainty exceeding noticeably the
experimental  one.  The recent aforesaid values of $S_{34}(0)$
recommended in Refs.[6-9] and \cite{Brown07} have nevertheless been
obtained from the analysis of the precisely measured data for
$S_{34}^{exp}(E)$ by means of the artificial renormalization of  the
energy dependence of the R-matrix calculation \cite{PD}   and of the
resonating-group method calculation \cite{Kaj86} for $S_{34}(E)$ to
the corresponding experimental data, respectively.

The theoretical calculations of   $S_{34}(0)$ performed within
different methods also show  considerable spread  [11-14]. The
 aforesaid resonating-group method calculations of  $S_{34}(0)$ performed in Ref.\cite{Kaj86} show
considerable sensitivity to the form of the effective NN interaction
used and the estimates have been obtained within the range of
0.312$\le S_{34}(0)\le$ 0.841 ${\rm  {keV\,\,\,b}}$. Calculations
performed in microscopic single-channel (($\alpha+^3{\rm {He}}$))
and two-channel (($^3{\rm {He}}+\alpha $) and ($p+^6Li$)) cluster
models gave the values of $S_{34}(0)$= 0.56 ${\rm  {keV\,\,\,b}}$
\cite{Lan86}( $S_{34}(0)$= 0.52 ${\rm  {keV\,\,\,b}}$ \cite{Cs00})
and of $S_{34}(0)$=0.83 ${\rm  {keV\,\,\,b}}$ \cite{Cs00},
respectively, that is, the estimate of the value of the $S_{34}(0)$
strongly changes when the model space is expanded. The calculations
performed by the authors of Refs.\cite{Moh93} and \cite{Dub95} in
the two-body potential model with
 different forms of the two-body potential gave the  values of
$S_{34}(0)$=0.516 and about of 0.5 ${\rm  {keV\,\,\,b}}$,
respectively, although different values of 1.174 in \cite{Moh93}
and 1.0 in \cite{Dub95} have been used for the spectroscopic
factor for the ($\alpha +^3{\rm {He}}$)-configuration in $^7{\rm
{Be}}$. Calculations performed in the variational Monte-Carlo
technique (VMCT) with seven-particle wave functions derived from
realistic NN interaction gave $S_{34}(0)\approx 0.40$ ${\rm
{keV\,\,\,b}}$ \cite{Nol01}.  But, as it was emphasized in paper
\cite{Nol01}, serious problems occur with the normalization for
the calculated astrophysical $S$-factor $S(E)$ in respect to  the
experimental data. The estimation of $S_{34}(0)$=0.52$\pm$ 0.03
${\rm {keV\,\,\,b}}$ \cite{Igam97} also should be  noted. The
latter has been obtained within the framework of the asymptotic
method developed in \cite{Igam95,Mukh0} based on the idea proposed
in paper \cite{Chris61}. This idea is based on the assumption
about the fact that low-energy direct radiative captures in light
nuclei (${\rm A}(a,\gamma ){\rm B}$) proceed mainly in regions
well outside the range of the internuclear interactions. But in
Ref. \cite{Igam97} the contribution  from the nuclear interior
($r<$ 4 fm) to the amplitude was assumed to be negligibly small.
In this assumption from the analysis of the experimental
astrophysical $S$-factors for the direct capture   $^3{\rm
{He}}(\alpha,\gamma)^7{\rm {Be}}$(g.s.) and $^3{\rm
{He}}(\alpha,\gamma)^7{\rm {Be}}$(0.429 MeV) reactions in the
energy range 180$\lesssim E \lesssim$ 500 keV \cite{Osb82} the
values of the nuclear vertex constants (NVC)  for the virtual
decays $^7{\rm {Be}}(g.s.)\to\alpha +^3{\rm {He}}$ and  $^7{\rm
{Be}}(0.429 MeV)\to\alpha +^3{\rm {He}}$ \cite{Blok77} (or the
respective asymptotic normalization coefficients (ANC) for $^3{\rm
{He}}+\alpha\to^7{\rm {Be}}$(g.s.) and $^3{\rm
{He}}+\alpha\to^7{\rm {Be}}$(0.429 MeV))  obtained were then used
for calculations of the astrophysical $S$-factors for the same
reactions at $E<$180 keV, including $E$=0. However,  the
experimental astrophysical $S$-factors for the direct capture
$^3{\rm {He}}(\alpha,\gamma)^7{\rm {Be}}$ reactions \cite{Osb82}
used in \cite{Igam97} for the analysis have considerable spread.
Consequently, the values of the ANC's $^3{\rm
{He}}+\alpha\to^7{\rm {Be}}$ and the $S_{34}(0)$ obtained in
\cite{Igam97} may not be enough accurate. Therefore, determination
of precise experimental values of the ANC's for $^3{\rm
{He}}+\alpha\to^7{\rm {Be}}$(g.s.) and $^3{\rm
{He}}+\alpha\to^7{\rm {Be}}$(0.429 MeV) is highly desirable since
it has direct effects in the correct extrapolation of the $^3{\rm
{He}}(\alpha,\gamma)^7{\rm {Be}}$ astrophysical $S$-factor at
solar energies \cite{Chris61,Igam07}.

In this work      new analysis of the highly precise experimental
astrophysical $S$-factors for the direct capture $^3{\rm
{He}}(\alpha,\gamma)^7{\rm {Be}}$ reaction at extremely low energies
($\gtrsim$ 90 keV)  [6-10] is performed within the modified two -
body potential approach \cite{Igam07} to obtain $^{\glqq}$indirectly
measured\grqq\, values both of the ANC's (the NVC's) for $^3{\rm
{He}}+\alpha\to^7{\rm {Be}}(g.s.)$ and $^3{\rm {He}}+\alpha\to^7{\rm
{Be}}(0.429)$, and of $S_{34}(E)$ at $E\le$ 90 keV, including $E$=0.
In the present work we show that one can extract ANC's for   $^3{\rm
{He}}+\alpha\to^7{\rm {Be}} $ directly from the $^3{\rm
{He}}(\alpha,\gamma)^7{\rm {Be}}$ reaction where the ambiguities
inherent for the standard two -body potential model calculation of
the $^3{\rm {He}}(\alpha,\gamma)^7{\rm {Be}}$ reaction  being
connected with the choice of the geometric parameters (the radius
$r_o$ and the diffuseness $a$) for the Woods--Saxon potential and
the spectroscopic factors, can be reduced in the physically
acceptable limit, being within the experimental errors for the
$S_{34}(E)$.

The contents of this paper are as follows. In Section 2   basic
formulae of the modified two-body potential approach to the direct
radiative capture $^3{\rm {He}}(\alpha,\gamma)^7{\rm {Be}}$
reaction are given. There the analysis of the precise measured
astrophysical $S$-factors for the direct radiative capture $^3{\rm
{He}}(\alpha,\gamma)^7{\rm {Be}}$ reaction is performed
(Subsections 2.2--2.4). The conclusion is given in Section 3.

 \section{Analysis of   $^3{\rm {He}}(\alpha,\gamma)^7{\rm {Be}}$ reaction}

\subsection{Basic formulae}

\hspace{0.7cm}  Here we give the   formulae specialized for the
$^3{\rm {He}}(\alpha,\gamma)^7{\rm {Be}}$   astrophysical
$S$-factor. Let us write    $l_f$ ($j_f$) for the relative orbital
(total) angular moment  of $^3{\rm {He}}$ and $\alpha$-particle in
nucleus $^7{\rm {Be}} (\alpha+^3{\rm {He}})$, $l_ i$ ($j_ i$) for
the orbital (total) angular moment  of the relative motion of the
colliding particles in the initial state, $\lambda $ for multipole
order of the electromagnetic transition, $\eta_f$($\eta_i$) for the
Coulomb parameter for the $^7{\rm {Be}}(=\alpha+^3{\rm {He}})$ bound
($^3{\rm {He}}\alpha$-scattering) state and $\mu$ for the reduced
mass of the ($^3{\rm {He}}\alpha$)-pair. For the $^3{\rm
{He}}(\alpha,\gamma)^7{\rm {Be}}$ reaction populating the ground and
first excited ($E^*$=0.429 MeV; $J^{\pi}$=1/2$^-$) states of $^7{\rm
{Be}}$, the values of $j_f$ are taken to be equal to 3/2 and 1/2,
respectively, the value of $l_f$ is taken to be equal to 1 as well
as   $l_i$=0, 2 for the $E1$-transition and $l_ i$=1 for the
$E2$-transition.

 According to \cite{MGTPHR01,Igam07}, for fixed $l_f$ and $j_f$ we can write the astrophysical
 $S$ -factor, $S_{l_fj_f}(E)$,   in the following form
\begin{equation} S_{l_fj_f}(E)=C^2_ {l_fj_f}{\cal
{R}}_{l_fj_f}(E,C_{l_fj_f}^{(sp)}). \label{12}
\end{equation}
Here, $C_ {l_fj_f}$ is   the ANC  for  $^3{\rm {He}}+\alpha\to^7{\rm
{Be}} $, which determines    the amplitude of the tail of the
$^7{\rm {Be}}$ nucleus  bound state wave function in the
($\alpha+^{3}He$)-channel and is  related to the NVC $G_{l_fj_f}$
for the virtual decay $^7{\rm {Be}}\to\alpha + ^3{\rm {He}}$ and  to
the spectroscopic factor $Z_{l_fj_f}$ for the ($^3{\rm
{He}}+\alpha$)-configuration with the quantum numbers $l_f$ and
$j_f$ in the $^7{\rm {Be}}$ nucleus as \cite{Blok77}
\begin{equation}
G_{l_fj_f}=-i^{l_f+\eta_f}\frac{\sqrt{\pi}}{\mu}C_{l_fj_f} \label{6}
\end{equation}
and
\begin{equation}
C_{l_fj_f}= Z_{l_fj_f}^{1/2}C^{(sp)}_{l_fj_f},\label{13a}
\end{equation}
respectively, and
\begin{equation}
{\cal {R}}_{l_fj_f}(E,C^{(sp)}_{l_fj_f})=\frac{
\tilde{S}_{l_fj_f}(E)}{(C^{(sp)}_{l_fj_f})^2}, \label{13}
\end{equation}
where
$\tilde{S}_{l_fj_f}(E)=\sum_{\lambda}\tilde{S}_{l_fj_f\lambda}(E)$
is the single-particle astrophysical $S$-factor \cite{An99} and
$C^{(sp)}_{l_fj_f}$ is the single-particle ANC, which determines
the amplitude of the tail of the single-particle wave function of
the bound $^7{\rm {Be}}$($\alpha+^3{\rm {He}}$) state. In
(\ref{6}) the factor taking into account the nucleon's identity
\cite{Blok77} is absorbed in the $C_{l_fj_f}$. The single-particle
bound state wave function, $\varphi_{l_fj_f}(r)$, is determined by
the solution of the radial Schr\"{o}dinger equation with the
phenomenological Woods--Saxon potential for the given quantum
numbers $n$ ($n$ is the nodes of $\varphi_{l_fj_f}(r)$), $l_f$ and
$j_f$ as well as geometric parameters of $r_o$ and $a$, and with
depth adjusted to fit the binding energy of the $^7{\rm {Be}}$
bound state with respect to the ($\alpha+^3{\rm {He}}$)-channel.
Note that   in Eq.(\ref{13}) the dependence of the function ${\cal
{R}}_{l_fj_f}(E,C^{(sp)}_{l_fj_f})$ on the free parameter
$C^{(sp)}_{l_fj_f}$ also enters  through the single-particle wave
function $\varphi_{l_fj_f}(r; C^{(sp)}_{
l_fj_f})(\equiv\varphi_{l_fj_f}(r))$ \cite{Gon82}, and the
single-particle ANC $C^{(sp)}_{l_fj_f}$  in turn is itself a
function of the geometric parameters of   $r_o$ and  $a$, i.e.,
$C^{(sp)}_{l_fj_f}=C^{(sp)}_{l_fj_f}(r_0,a)$.

According to \cite{Igam07}, the peripheral character for the direct
capture $^3{\rm {He}}(\alpha,\gamma)^7{\rm {Be}}$ reaction is
conditioned by
\begin{equation} {\cal {R}}_{l_fj_f}(E,C^{(sp)}_{l_fj_f})=f(E)  \label{15}
\end{equation}
as a function of the $C_{l_fj_f}^{(sp)}$   within the energy range
$E_{min}\leq E \leq E_{max}$, where the left hand side (l.h.s.)  of
Eq.(\ref{15}) must not depend on $C_{l_fj_f}^{(sp)}$ for each fixed
$E$ from the aforesaid energy range, and by
\begin{equation}
C_{l_fj_f}^2=\frac{S_{l_fj_f}(E)}{{\cal{R}}_{l_fj_f}(E,C_{l_fj_f}^{(sp)})}
=const \label{16}
\end{equation}
for each fixed  $E$ and the function of ${\cal
{R}}_{l_fj_f}(E,C_{l_fj_f}^{(sp)})$  from (\ref{15}).

As  it was previously shown  in \cite{Igam07} and \cite{Igam08} for
the direct capture $t(\alpha,\gamma )^7{\rm {Li}} $ and ${\rm
{^7Be}}(p,\gamma){\rm {^8B}}$ reactions, respectively, fulfillment
of the conditions (\ref{15}) and (\ref{16}) enables one also to
obtain valuable information about the experimental
($^{\glqq}$indirectly measured\grqq\,) value of the ANC
$(C_{l_fj_f}^{exp})^2$ for $^3{\rm {He}}+\alpha\to^7{\rm {Be}} $ by
using $S^{exp}_{l_fj_f}(E)$ instead of $S_{l_fj_f}(E)$ in the right
hand side (r.h.s.) of Eq.(\ref{16}):
\begin{equation}
(C_{{l_fj_f}}^{exp})^2=\frac{S^{exp}_{l_fj_f}(E)} {{\cal
{R}}_{l_fj_f}(E,C_{l_fj_f}^{(sp)})}.
  \label{17} \end{equation}
Then the value of the ANC, $(C_{l_fj_f}^{exp})^2$, obtained from
Eq.(\ref{17})  together with the condition (\ref{15}) can be used
for calculation of $S_{l_fj_f}(E)$ at    energies of $E<E_{min}$ by
using  the following expression:
\begin{equation} S_{l_fj_f}(E)=(C_{l_fj_f}^{exp})^2{\cal {R}}_{l_fj_f}(E,C^{(sp)}_{l_fj_f}).
\label{18}
\end{equation}

Note  that the total astrophysical $S$-factor for the $^3{\rm
{He}}(\alpha,\gamma)^7{\rm {Be}}$(g.s.+0.429MeV) reaction is given
by
\begin{equation}
S_{34}(E)=\sum_{j_f=1/2,3/2} S_{l_fj_f}(E)= C_{1\,\,1/2}^2{{\cal
R}}_{1\,\,1/2}(E,C_{1\,\,1/2}^{(sp)})+C_{1\,\,3/2}^2{{\cal
R}}_{1\,\,3/2}(E,C_{1\,\,3/2}^{(sp)})\label{18a}
\end{equation}
\begin{equation}
=C_{1\,\,3/2}^2{{\cal R}}_{1\,\,3/2}(E,C_{1\,\,3/2}^{(sp)})[1+R(E)]
\label{18aa}
\end{equation}
\begin{equation}
=C_{1\,\,1/2}^2{{\cal
R}}_{1\,\,1/2}(E,C_{1\,\,1/2}^{(sp)})[1+R^{-1}(E)]\label{18ab}
\end{equation}
\begin{equation}
=C_{1\,\,3/2}^2[{{\cal
R}}_{1\,\,3/2}(E,C_{1\,\,3/2}^{(sp)})+\lambda_C{{\cal
R}}_{1\,\,1/2}(E,C_{1\,\,1/2}^{(sp)})],\label{18ac}
\end{equation}
where $R(E)=S_{1\,\,1/2}(E)/S_{1\,\,3/2}(E)=C_{1\,\,1/2}^2{{\cal
R}}_{1\,\,1/2}(E,C_{1\,\,1/2}^{(sp)})/C_{1\,\,3/2}^2{\cal
{R}}_{1\,\,3/2}(E,C_{1\,\,3/2}^{(sp)})$ is a branching ratio and
$\lambda_C=(C_{1\,\,1/2}/C_{1\,\,3/2})^2$.

Values obtained in such a way for the  $(C_{l_fj_f}^{exp})^2$ and
$S_{l_fj_f}(E)$ at   energies of $E<E_{min}$ can be considered as an
$^{\glqq}$indirect measurement\grqq\, of the ANC (or NVC) for
$^3{\rm {He}}+\alpha\to^7{\rm {Be}} $ and of the astrophysical
$S$-factor for the direct capture $^3{\rm {He}}(\alpha,\gamma)^7{\rm
{Be}}$ reaction at $E<E_{min}$, including $E=0$. It should be noted
that the expressions (\ref{12}) and (\ref{13})--(\ref{18ac}) allow
one to determine both the absolute value of ANC (or NVC) for $^3{\rm
{He}}+\alpha\to^7{\rm {Be}} $ and that of the astrophysical
$S$-factor $S_{l_fj_f}(E)$ for the peripheral direct capture $^3{\rm
{He}}(\alpha,\gamma)^7{\rm {Be}}$ reactions at extremely low
experimentally inaccessible  energy regions by means of the analysis
of the same precisely measured values of the experimental
astrophysical $S$-factors, $S^{exp}_{l_fj_f}(E)$ and
$S^{exp}_{34}(E)$.

\subsection{The asymptotic normalization coefficients for
 $^3{\rm {He}}+\alpha\to^7{\rm {Be}}$}

\hspace{0.7cm}   To determine the ANC values for the $^3{\rm
{He}}+\alpha\to^7{\rm {Be}}$(g.s)  and $^3{\rm {He}}+\alpha\to^7{\rm
{Be}}$(0.429 MeV) the experimental astrophysical $S$-factors,
$S^{exp}_{l_fj_f}(E)$, for the $^3{\rm {He}}(\alpha,\gamma)^7{\rm
{Be}}$   reaction populating the ground ($l_f=1$ and $j_f$=3/2) and
first excited ($E^*$=0.429 $MeV$; $J^{\pi }=1/2^-$,\,\,\,$l_f=1$ and
$j_f$=1/2) states are reanalyzed based on the   relation (\ref{12}),
the conditions (\ref{15}) and (\ref{16}), and   the relations
(\ref{17}) and (\ref{18}).  As it was mentioned above, the
experimental data have been obtained by different authors, which
have considerable spread with experimental uncertainty being   more
than 10\%. Recently, S.B. Nara Singh et al. \cite{Nara04}, D.
Bemmerer et al. \cite{Bem06}, Gy. Gy\"{u}ky et al. \cite{Gy07},  F.
Confortola et al. \cite{Con07} and   T.A.D. Brown et al.
\cite{Brown07} have apparently performed the most accurate direct
measurement of the total astrophysical $S$-factor for the $^3{\rm
{He}}(\alpha,\gamma)^7{\rm {Be}}$ reaction, covering the energy
ranges $E$=92.9--168.9 keV [7--9], 420--951 keV \cite{Nara04} and
   327--1235 keV \cite{Brown07} with absolute
uncertainty not exceeding 5\%. However, one should note  that the
experimental astrophysical $S$-factors for the the $^3{\rm
{He}}(\alpha,\gamma)^7{\rm {Be}}$ reaction populating to the first
and excited states of the residual nucleus have been separated only
for the energies of $E$=92.9, 105.6 and 147.7 keV in \cite{Con07}
and for all experimental points of $E$ from  the aforesaid energy
region in \cite{Brown07}. Whereas, in \cite{Brown07} the
experimental astrophysical $S$-factors  were measured by using two
different experimental approaches:the detection of the delayed
$\gamma$ ray from $^7{\rm {Be}}$ (the activation) and the
measurement of the prompt $\gamma$ emission (the prompt). So, in our
analysis we naturally use $S^{exp}_{34}(E)$ most recently
independently measured in Refs.[6--9] and \cite{Brown07}, since the
reaction under consideration is nonresonant   and, consequently,
proceeds mainly in regions well outside the internuclear interaction
range \cite{Chris61}.

 The Woods--Saxon potential split with a parity ($l$ - dependence) for the
spin-orbital term proposed by the authors of Refs. [28--30] is used
here for the calculations  of both   bound state  radial wave
function $\varphi _{l_fj_f}(r)$ and   scattering wave function
$\psi_{l_ij_i}(r)$.  It should be emphasized that the choice of this
potential is based on the following considerations. Firstly, this
potential form is justified from the microscopic point of view
because it makes it possible to take into account the Pauli
principle between nucleons in $^3{\rm {He}}$- and $\alpha$-clusters
in the ($\alpha+^3{\rm {He}}$) bound  state by means of inclusion of
deeply bound states forbidden by the Pauli exclusion principle, i.e.
without an explicit introduction of a repulsive core at  small
distance.  The latter  imitates  the additional node ($n$) arising
in the wave functions of $\alpha-^3{\rm {He}}$ relative motion  in
$^7{\rm {Be}}$. Secondly, this potential describes well  the phase
shifts for $ ^3{\rm {He}}\alpha$-scattering in the wide energy range
\cite{Neu75,Neu83}.

The test of the peripheral character of the $^3{\rm
{He}}(\alpha,\gamma)^7{\rm {Be}}$  reaction for the aforesaid
energy range has been made by means of verifying the conditions
(\ref{15}) and (\ref{16}), and by changing the geometric
parameters (radius $r_o$ and diffuseness $a$) of the adopted
Woods--Saxon potential using the procedure of the depth adjusted
to fit the binding energies , as it was done in Ref.
\cite{Igam07}.   According to Ref. \cite{Igam07}, we vary $r_o$
and $a$ in the physically acceptable ranges ($r_o$ in 1.62--1.98
fm and $a$ in 0.63--0.77 fm) in respect to the standard values
($r_o$=1.80 fm and $a$=0.70 fm \cite{Neu75,Neu83}). Such a choice
of the $r_o$ and $a$ parameters variation limit       allows us to
provide fulfillment of the conditions (\ref{15}) and (\ref{16}) in
the aforesaid energy range within the experimental errors for the
$S^{exp}_{l_fj_f}(E)$.

\begin{figure}[!h]
\epsfxsize=15.cm \centerline{\epsfbox{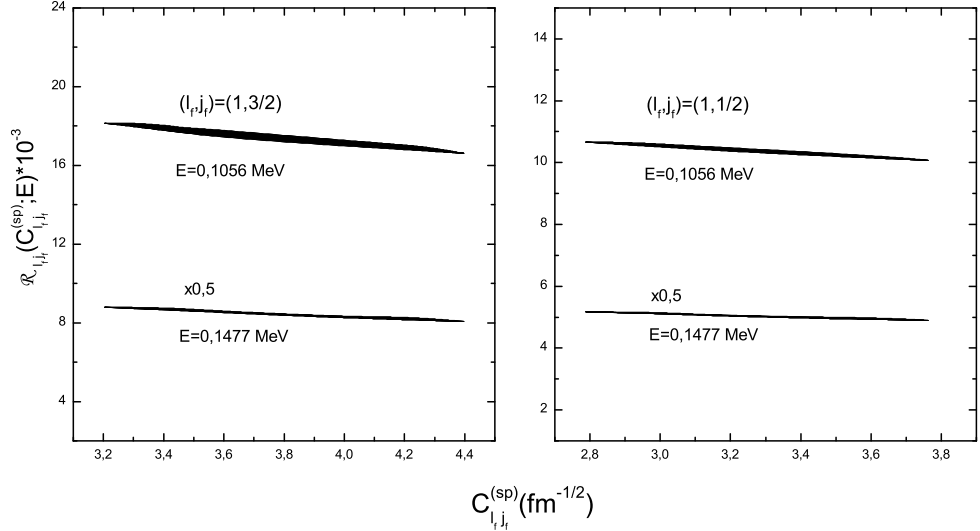}}
\caption{\label{fig1} The dependence of ${\cal
{R}}_{l_fj_f}(E,C^{(sp)}_{l_fj_f})$ as a function of the
single-particle ANC, $C^{(sp)}_{l_fj_f}$, for  the $^3{\rm
{He}}(\alpha,\gamma)^7{\rm {Be}}$(g.s.) ($(l_f,j_f)$=(1,3/2))  and
 $^3{\rm {He}}(\alpha,\gamma)^7{\rm {Be}}$(0.429
MeV  ($(l_f,j_f)$=(1,1/2)) reactions at  different energies E.}
\end{figure}
\begin{figure}[!h]
\epsfxsize=12.cm \centerline{\epsfbox{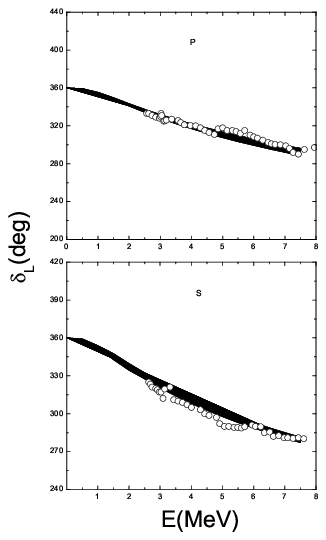}}
\caption{\label{fig2}The energy dependence of the $^3{\rm
{He}}\alpha$-elastic scattering phase
 shifts for different partial waves. The experimental
data are from  \cite{Barn64}. The bands are our calculated data.
The width of the bands for fixed energies corresponds to the
variation of the parameters $r_o$ and $a$ of the adopted
Woods--Saxon potential within the intervals of $r_o$=1.62 to 1.98
fm and $a$=0.63 to  0.77 fm.}
\end{figure}

As an illustration,  Fig.\ref{fig1} shows  plots of the ${\cal
{R}}_{l_fj_f}(E,C^{(sp)}_{l_fj_f})$  dependence on the
single-particle ANC, $C^{(sp)}_{l_fj_f}$  for $l_f$= 1 and
$j_f$=3/2 and 1/2 only for  the two  values of   energy  $E$. The
width of the band for these curves  is the result of the weak
$^{\glqq}$residual\grqq\,$(r_o,a)$-dependence of the ${\cal
{R}}_{l_fj_f}(E,C^{(sp)}_{l_fj_f})$  on the parameters $r_o$ and
$a$ (up to $\pm 2\%$) for the
$C^{(sp)}_{l_fj_f}=C^{(sp)}_{l_fj_f}(r_o,a)=const$
\cite{Gon82,Igam07}. The same dependence is also observed at other
energies.   It is seen that for the calculated values of   ${\cal
{R}}_{l_fj_f}(E,C^{(sp)}_{l_fj_f})$  the  dependence on the
$C^{(sp)}_{l_fj_f}$ values  is  rather weak (no more than  $\pm$
5.0\% )  in the interval of $3.205\leq C^{(sp)}_{13/2}\leq 4.397$
fm$^{-1/2}$ ($2.788\leq C^{(sp)}_{11/2}\leq 3.763$ fm$^{-1/2}$)
for the $^3{\rm {He}}(\alpha,\gamma)^7{\rm {Be}}$(g.s.) ($^3{\rm
{He}}(\alpha,\gamma)^7{\rm {Be}}$(0.429 MeV)) reaction, which
corresponds to the parameters of the adopted Woods--Saxon
potential $r_o$ ranging from 1.62--1.98 fm and $a$ in the range of
0.63--0.77 fm.  It follows from here that the condition (\ref{15})
is satisfied for the considered reaction within the uncertainties
not exceeding the experimental errors of $S^{exp}_{l_fj_f}(E)$.

We also calculated  the $^3{\rm {He}}\alpha$-elastic scattering
phase shifts by variation of the parameters $r_o$ and $a$ in the
same range for the adopted Woods--Saxon potential. The results of
the calculations corresponding to $s$- and $p$-waves are presented
in Fig.\ref{fig2} in which the width of the bands corresponds to a
change of phase shifts values  with respect to  variation of
values of the $r_o$ and $a$ parameters.  As it is seen from
Fig.\ref{fig2}, the experimental phase shifts \cite{Barn64}   are
well reproduced within uncertainty of about $\pm$ 5\%.

\begin{figure}[!h]
\epsfxsize=15.cm \centerline{\epsfbox{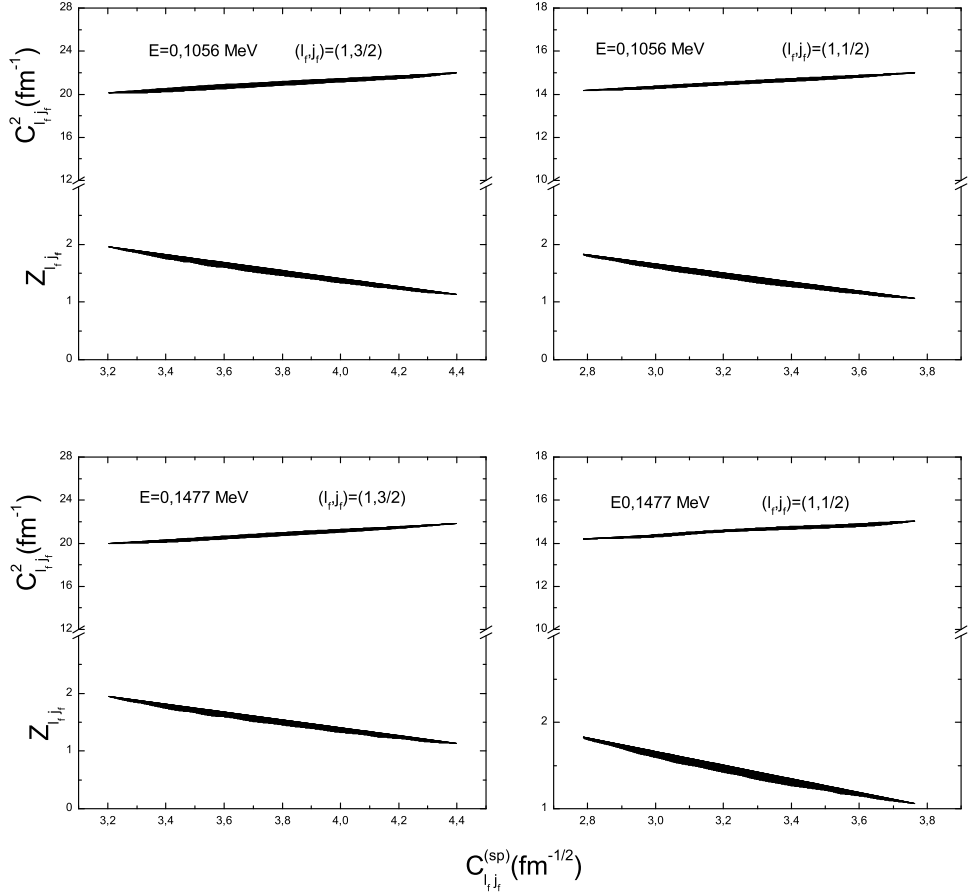}}
\caption{\label{fig3}The dependence of the ANC's $C_{l_fj_f}$
(upper band) and the spectroscopic factors $Z_{l_fj_f}$ (lower
band) on the single-particle ANC $C^{(sp)}_{l_fj_f}$  for the
$^3{\rm {He}}(\alpha,\gamma)^7{\rm {Be}}$(g.s.) (the left column,$
(l_f,j_f)$=(1,3/2)) and $^3{\rm {He}}(\alpha,\gamma)^7{\rm
{Be}}$(0.429 MeV) (the right column, $(l_f,j_f)$=(1,1/2))
reactions at  different energies E.}
\end{figure}

This circumstance allows us to test the condition (\ref{16}), which
is no less essential for the peripheral character of these
reactions, at the energies of $E$= 92.9, 105.6 and 147.7 keV for
which the $^3{\rm {He}}(\alpha,\gamma)^7{\rm {Be}}$(g.s.) and
$^3{\rm {He}}(\alpha,\gamma)^7{\rm {Be}}$(0.429 MeV) astrophysical
$S$-factors were separately measured  in  \cite{Con07}. As an
illustration,  for the same energies $E$  as in Fig.\ref{fig1} we
present in Fig.\ref{fig3} (the upper panels) the results of
$C_{l_fj_f}^2$-value calculation given by Eq.(\ref{16})
($(l_fj_f)$=(1$\,\,$ 3/2) and (1$\,\,$ 1/2) ) in which instead of
the $S_{l_fj_f}(E)$ the experimental $S$- factors for the $^3{\rm
{He}}(\alpha,\gamma)^7{\rm {Be}} $ reaction populating the ground
and first excited states of $^7{\rm {Be}}$ were taken. It is also
noted that the same dependence occurs for other considered energies.
It is seen from this figure that the obtained $C_{l_fj_f}^2$ values
also weakly depend  on the $C^{(sp)}_{l_fj_f}$-value. However, the
values of the spectroscopic factors  $Z_{1\,\,3/2}$ and
$Z_{1\,\,1/2} $ corresponding to the $(\alpha +^3{\rm
{He}})$-configuration for $^7{\rm {Be}}(g.s.)$ and $^7{\rm
{Be}}(0.429 keV)$, respectively, change  strongly (see, the lower
panels in Fig.\ref{fig3}).
The calculation shows that  the uncertainty in ${\cal
{R}}_{l_fj_f}(E,C^{(sp)}_{l_fj_f})$ and $C_{l_fj_f}^2$ is up to
$\pm 5.0$\% relative to the central values of ${\cal
{R}}_{l_fj_f}(E,C^{(sp)}_{l_fj_f})$ and $C_{l_fj_f}^2$, obtained
for the standard values of $r_o=1.80$  fm  and $a=0.60$
 fm,  for the ($r_o$, $a$)-pair varying in the above mentioned
intervals for $r_o$ and $a$, while the uncertainty in the
$Z_{l_fj_f}$   is about $\pm 30$\%. It should be noted that the
uncertainty in  the $C_{l_fj_f}^2$ values becomes even less when in
(\ref{16}) one uses the experimental astrophysical $S$-factors
corresponding to smaller energies. Thus, the peripheral character of
the reactions under consideration allows one to determine the
$C_{1\,\,3/2}^2$  and $C_{1\,\,1/2}^2$  values for the  $ \alpha
+^3{\rm {He}} \rightarrow ^7{\rm {Be}}(g.s.)$ and $\alpha +^3{\rm
{He}}\rightarrow ^7{\rm {Be}}$(0.429 keV), respectively, with a
maximal uncertainty of about $\pm$  5.0\% when the geometric
parameters $r_o$ and $a$ are varied within the aforesaid ranges and
the experimental data  are used at the aforesaid three energies in
the  analysis.

\begin{figure}[!h]
\epsfxsize=15.cm \centerline{\epsfbox{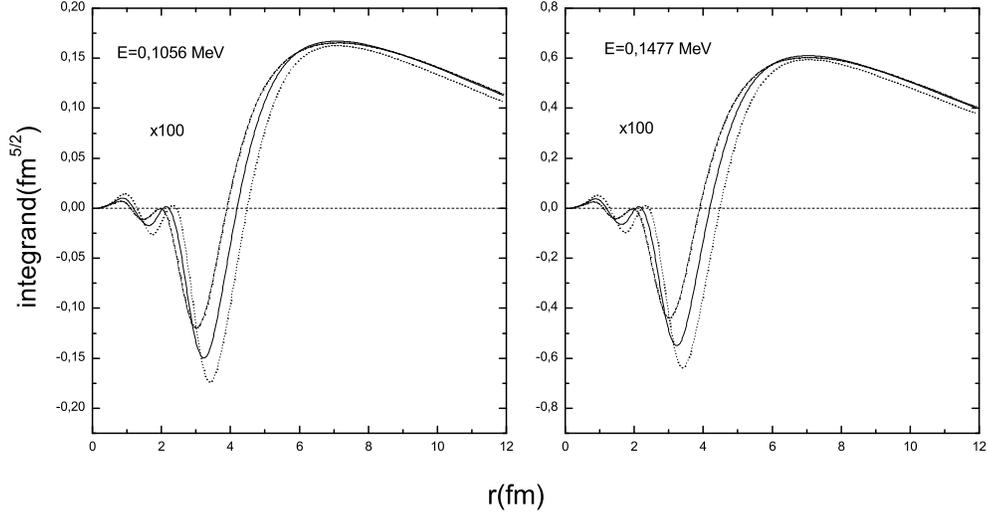}}
\caption{\label{fig4}  The integrand of the radial integral (10)
of Ref.\cite{Igam07} for the $^3{\rm {He}}(\alpha,\gamma)^7{\rm
{Be}}$(g.s.) reaction at energies $E$=0.1056 and 0.506 and 0.951
MeV for different sets  of ($r_o;a$)-pairs:(1.62 fm;0.63 fm)
(dashed line), (1.80 fm;0.70 fm) (solid line) and (1.98 fm;0.77
fm) (dotted line).}
\end{figure}

For different energies $E$ we also estimate a relative
contribution of the nuclear interior ($r<r_N$) to the
astrophysical $S$-factors for the  $^3{\rm
{He}}(\alpha,\gamma)^7{\rm {Be}}$  reaction populating the ground
and first excited states in dependence on the variation
$C_{l_fj_f}^{(sp)}$ (or $r_o$ and $a$) introducing  the cutoff
radius $r_{cut}$ ($r_{cut}\approx r_N$) in the lower limit of
integration of the radial integral (10) of Ref.\cite{Igam07}
entering in the amplitude of the reaction under consideration .
With this aim one considers the ratio
$\Delta(E,C^{(sp)}_{l_fj_f};r_{cut})= \mid {\cal
{R}}_{l_fj_f}(E,C^{(sp)}_{l_fj_f})-\tilde {\cal
{R}}_{l_fj_f}(E,C^{(sp)}_{l_fj_f};r_{cut})\mid /{\cal
{R}}_{l_fj_f}(E,C^{(sp)}_{l_fj_f})$, where $\tilde {\cal
{R}}_{l_fj_f}(E,C^{(sp)}_{l_fj_f};r_{cut})$ is given by Eqs.(10)
and (13) of Ref.\cite{Igam07} but in the radial integral (10) of
Ref.\cite{Igam07} the integration over $r$ is performed in the
interval $r_{cut}\le r\le\infty$, i.e. $\tilde {\cal
{R}}_{l_fj_f}(E,C^{(sp)}_{l_fj_f};0)={\cal
{R}}_{l_fj_f}(E,C^{(sp)}_{l_fj_f})$. The ${\cal
{R}}_{l_fj_f}(E,C^{(sp)}_{l_fj_f})$ and $\tilde {\cal
{R}}_{l_fj_f}(E,C^{(sp)}_{l_fj_f};r_{cut})$ functions were
calculated for different values of the single-particle ANC
$C^{(sp)}_{l_fj_f}$ ( or the parameters $r_o$ and $a$). A value of
the cutoff radius is taken as in Ref. \cite{Rolfs79}, that is
$r_{cut}= r^o_{\alpha t}= 1.36(4^{1/3}+3^{1/3})$=4.12 fm, as well
as $r_{cut}$=4.00 fm and 4.25 fm.  The calculation   of    $\Delta
(E,C^{(sp)}_{l_fj_f};r_{cut})$ performed   at  different energies
shows that the quantities of $\Delta(E,C^{(sp)}_{l_fj_f};r_{cut})$
change from 5.4\% up to 15.5\% under variation of
$C^{(sp)}_{l_fj_f}$ and $r_{cut}$. The calculation  shows that the
contribution of the nuclear interior ($r<r_N$) to the
astrophysical $S$-factors calculated for different sets of
geometric parameters $r_o$ and $a$ of the Woods--Saxon potential,
and values of $r_{cut}$ does not exceed about 15.5\% and this
small quantity is due mainly to oscillations observed in the
integrand of the radial integral (10) of Ref.\cite{Igam07}. As an
illustration of this fact, in Fig.\ref{fig4} we show a dependence
of the integrand of the radial integral (10) of Ref.\cite{Igam07}
on geometric parameters $r_o$ and $a$ of the Woods--Saxon
potential and values of $r_{cut}$ for the $^3{\rm
{He}}(\alpha,\gamma)^7{\rm {Be}}$(g.s.) reaction at different
energies.
As one can see  from Fig.\ref{fig4} the integrand of the radial
integral  changes with the variation of the geometric parameters
$r_o$ and $a$, which is associated   with changes of the
calculated bound ($\alpha+^3{\rm {He}}$) state wave function and
the calculated $^3{\rm {He}}\alpha$-scattering wave function, and
these wave functions indeed reached simultaneously their
asymptotic form for $r\gtrsim$ 5.0 fm. Such a change leads to
calculated $\tilde {S}_{l_fj_f}(E)$ that vary by 1.75 times over
the energy region 92.9$\le E\le$ 1200 keV, while the calculated
values of the function ${\cal {R}}_{l_fj_f}(E,C^{(sp)}_{l_fj_f})$
change by only $\pm 5\%$ with respect to  the value of ${\cal
{R}}_{l_fj_f}(E,C^{(sp)}_{l_fj_f})$ corresponding to the standard
values of $r_o$=1.80 fm and $a$=0.70 fm. Besides,  it is seen from
Fig.\ref{fig4}    that the behavior of the integrand in the radial
integral (10) of Ref.\cite{Igam07} over a wide energy range
provides a strong suppression of the contribution only from the
part of the nuclear interior with $0\le r\lesssim 2.0$ fm to the
integral (10) of Ref.\cite{Igam07}. But a noticeable change of the
integrand of the aforesaid integral is observed with the variation
of the parameters $r_o$ and $a$ in the range $2.0\lesssim
r\lesssim 6.0$ fm. The similar situation occurs for the  $ ^3{\rm
{He}}(\alpha,\gamma)^7{\rm {Be}}$(0.429 MeV) reaction. Therefore,
the choice of the cutoff radius $r_{cut}$ becomes ambiguous since
a fitted value of $r_{cut}$ also becomes dependent  on the
parameters $r_o$ and $a$.  In this connection one would like to
note the following. In paper \cite{Igam97} the calculation of the
astrophysical $S$-factors for the reactions under consideration
has been carried out using the expression (\ref{12}) but
introducing the cutoff radius $r_{cut}$ in the lower limit of
integration in the radial integral (10) of Ref.\cite{Igam07} and
replacing   the bound state wave function $\varphi_{l_fj_f}(r)$
with its asymptotic form starting from  $r=r_{cut}$. At this the
best fitting of the calculated $S_{l_fj_f}(E))$ to the
experimental ones \cite{Osb82} was reached when the cutoff radius
was $r_{cut}$=4.0 fm.  It is seen from here that in paper
\cite{Igam97}  the contribution of the nuclear interior to the
calculated astrophysical $S$-factors was indeed
underestimated\footnote{It should be noted that there is a
misprint in the line 36 upper of section 3 of \cite{Igam97}. There
the phrase "no more than 1\% to" must be written as  "no more than
10\% to".}, since   contribution of the nuclear interior $0< r\le
4.0$ fm to the calculated astrophysical $S$-factors, which   is up
to about 14\%, has not been taken into account.  Here, firstly,
the contribution of the nuclear interior ($r\le r_N$) to the
calculated astrophysical $S$-factors is taken into account in a
correct way by means of the appropriate choice of the adopted
potential both for the initial state and for final state of the
reactions under consideration. Secondly, the problem of the
ambiguity connected with the strong ($r_o,a$)-dependence of the
calculated astrophysical $S$-factors is removed by inclusion of
the information about ANC (or NVC). The latter reduces this
ambiguity to minimum. At last, in the present work the more
precise experimental data for the $^3{\rm
{He}}(\alpha,\gamma)^7{\rm {Be}}$ astrophysical $S$-factors
[6--10] than those in \cite{Igam97} are used for the analysis.

Thus the scrupulous analysis performed here quantitatively shows
that the $^3{\rm {He}}(\alpha,\gamma)^7{\rm {Be}}$ reaction within
the considered energy ranges is peripheral.

\begin{figure}[!h]
\epsfxsize=13.0 cm \centerline{\epsfbox{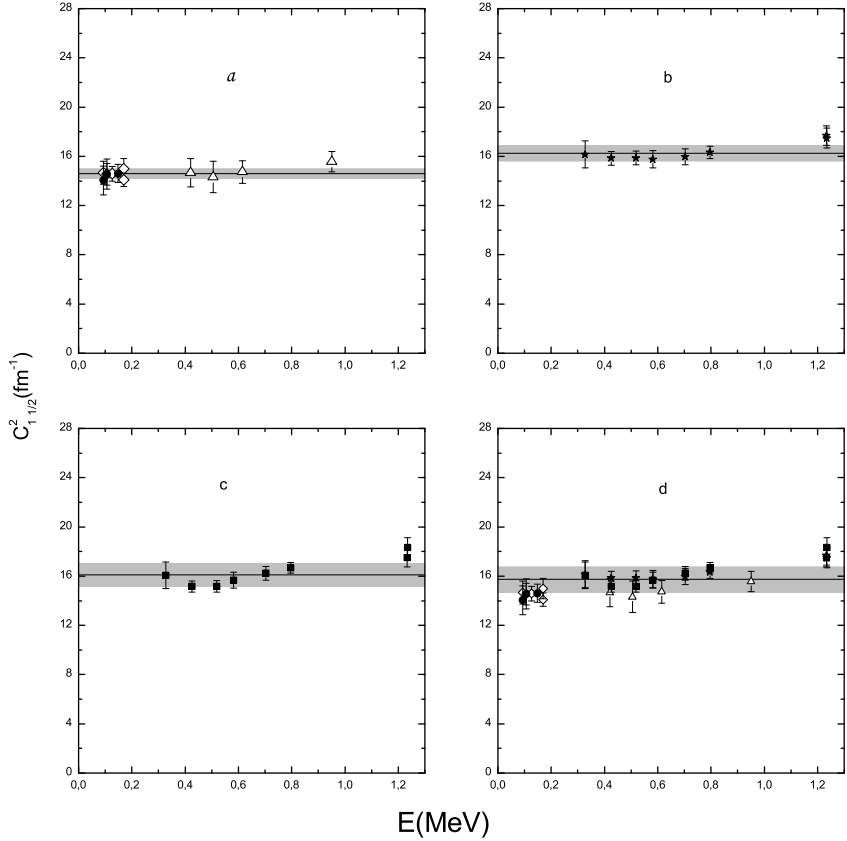}}
\caption{\label{fig5}The  values of the ANC's, $C^2_{1\,\,3/2}$,
for the $\alpha +^3{\rm {He}}\to^7{\rm {Be}}$(g.s.) for each
energy $E$ experimental point.     The opened triangle  and
diamond symbols (filled star and square symbols) are  data
obtained  by using the total (separated)  experimental
astrophysical $S$-factors from
 \cite{Nara04} and  [7--9] ($a$)( from  \cite{Brown07}, the
activation ($b$) and the prompt ($c$)), respectively, while
   filled circle symbols are data obtained from the separated experimental astrophysical $S$-factors from
  Refs.[7--9]. The symbols in
 ($d$)  are data obtained from all experimental astrophysical $S$-factors. The solid lines present our
results for the weighted means. Everywhere the width of each of
the band is the weighted uncertainty.}
\end{figure}

\begin{figure}[!h]
\epsfxsize=13.cm \centerline{\epsfbox{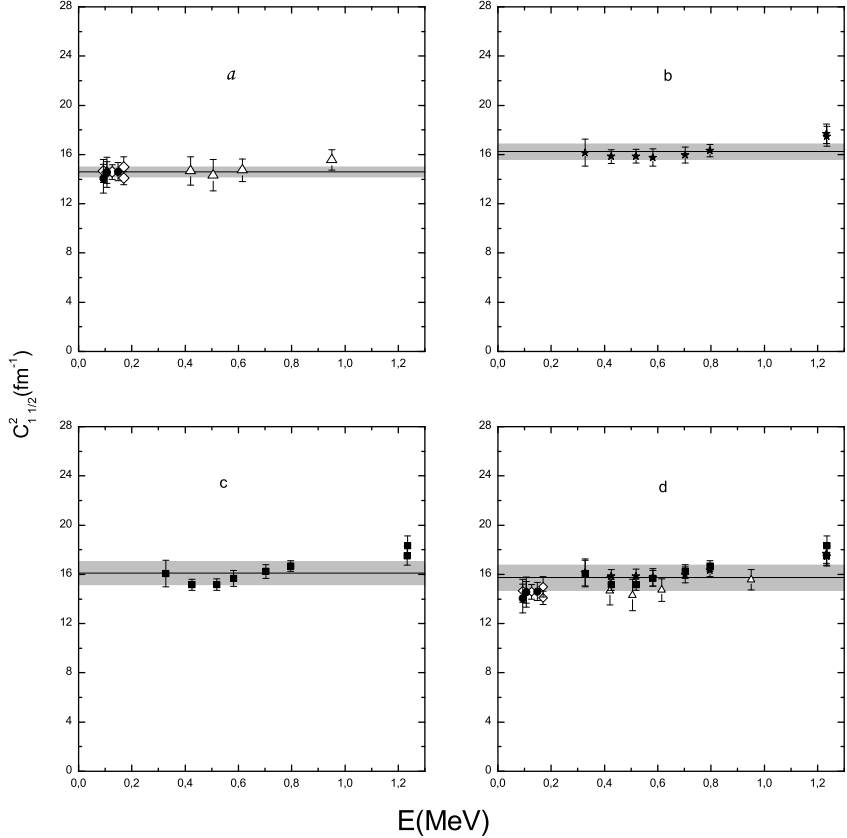}}
\caption{\label{fig6}The same as Fig.\ref{fig5} for the $\alpha
+^3{\rm {He}}\to^7{\rm {Be}}$(0.429 MeV).}
\end{figure}

For  each  energy $E$  experimental point ($E$=92.9, 105.6 and 147.7
keV) the values of the ANC's are obtained for the $\alpha + ^3{\rm
{He}}\rightarrow^7{\rm {Be}}(g.s.)$ and $\alpha +^3{\rm
{He}}\rightarrow ^7{\rm {Be}}$(0.429 MeV) by using the corresponding
experimental astrophysical $S$-factor ($S_{1\,\,3/2}^{exp}(E)$ and
$S_{1\,\,1/2}^{exp}(E)$, (the activation)) \cite{Con07} in the ratio
of the r.h.s. of the relation (\ref{16}) instead of the
$S_{l_fj_f}(E)$ and the central values of ${\cal
{R}}_{l_fj_f}(E,C^{(sp)}_{l_fj_f})$ corresponding to the adopted
values of the parameters $r_{0}$ and $a$.    The results of the
ANC's, $(C_{1\,\,3/2}^{exp})^2$ and $(C_{1\,\,1/2}^{exp})^2$  for
these three energy $E$ experimental points are displayed in
Figs.\ref{fig5}\emph{a} and \ref{fig6}\emph{a} (filled circle
symbols), and the second and third columns of Table \ref{table1}.
The uncertainties pointed in this figure correspond to those found
from (\ref{16}) (averaged square errors (a.s.e.)), which include the
total experimental errors in the corresponding experimental
astrophysical $S$-factor and the aforesaid uncertainty in the ${\cal
{R}}_{l_fj_f}(E,C^{(sp)}_{l_fj_f})$. One should note that the same
results for the ANC's, $C_{1\,\,3/2}^2$ and $C_{1\,\,1/2}^2$, (the
opened symbols in Figs.\ref{fig5}\emph{a} and \ref{fig6}\emph{a})
are obtained when $S^{exp}_{34}(E)$ (or $S^{exp}_{34}(E)$ and
$R^{exp}(E)$) \cite{Bem06,Con07}  are used in Eq.(\ref{18a}) (or in
Eq.(\ref{18aa}) and (\ref{18ab})) instead of $S_{34}(E)$
($S_{34}(E)$ and $R(E)$).
Then in Eq.(\ref{18ac}), inserting the weighted means of
$\lambda_C$ ($\lambda_C$=0.666), obtained from the three data (the
filled circle symbols) plotted in Figs.\ref{fig5}\emph{a} and
\ref{fig6}\emph{a}, and replacing  of the $S _{34}(E)$ in the
l.h.s. of Eq.(\ref{18ac}) with $S^{exp}_{34}(E)$ for the others,
the two experimental points  of energy $E$ ($E$=126.5 and 168.9
keV) from \cite{Bem06}, four  one $E$ ($E$=420.0, 506.0, 615.0 and
951.0 keV) from \cite{Nara04} and the three one $E$ ($E$=93.3,
106.1 and 170.1 keV) from \cite{Con07} can also determine values
of ANC's, $C_{1\,\,3/2}^2$ and $C_{1\,\,1/2}^2$. The results of
the ANC's are also displayed in Figs.\ref{fig5}\emph{a} and
\ref{fig6}\emph{a} (the   opened diamond and triangle symbols) and
the second and third columns of Table \ref{table1}. The same way
the values of the ANC's ($C_{1\,\,3/2}^2$ and $C_{1\,\,1/2}^2$)
are obtained by using the separated experimental astrophysical
$S$-factors ($S_{1\,\,3/2}^{exp}$ and $S_{1\,\,1/2}^{exp}$)
\cite{Brown07} obtained by using the experimental  astrophysical
$S$-factors (the activation ($b$) and the prompt ($c$)). The
results for ANC's are presented in Figs.\ref{fig5}\emph{b} (the
activation) and \ref{fig5}\emph{c} (the prompt) as well as
Figs.\ref{fig6}\emph{b} and \ref{fig6}\emph{c}. The weighted means
of the ANC--values and their uncertainties, deduced separately
from each experimental data, are displayed by the solid lines and
the band widths, respectively, in these figures and also presented
in the second and fourth columns of Table \ref{table2}. The
corresponding NVC-values are also presented in the third and fifth
columns of Table \ref{table2}. Besides, in Figs.\ref{fig5}\emph{d}
and \ref{fig6}\emph{d} (the solid line) as well as the second and
third columns of Table \ref{table2}, the weighted means of the
ANC--values derived from all of  the experimental points of
Figs.\ref{fig5} (\emph{a}, \emph{b} and \emph{c}) and
Figs.\ref{fig6} (\emph{a}, \emph{b} and \emph{c}) are also
presented.
 As it is also seen from Figs.\ref{fig5}  and \ref{fig6} (the second and
third (the second and fourth ) columns of Table \ref{table1} (Table
\ref{table2})) the values of the ANC's, obtained from the analysis
of the experimental astrophysical $S$-factors measured by the
authors of Refs.[6--10] in different intervals of energy $E$, agree
rather well with each other within about 12\%. Also, it is seen from
here that the ratio in the r.h.s. of the relation (\ref{17})  does
not practically depend on the energy $E$, although absolute values
of the corresponding experimental astrophysical $S$-factors for the
reactions under consideration depend noticeably on the energy and
change by up to about 1.7 times when $E$ changes from 92.6 keV to
1200 keV.

This fact allows us to conclude that the energy dependence of the
experimental astrophysical $S$-factors [6--10] is well determined by
the calculated function ${\cal {R}}_{l_fj_f}(E,C^{(sp)}_{l_fj_f})$
and ${{\cal R}}_{13/2}(E,C_{13/2}^{(sp)})+\lambda_C{{\cal
R}}_{11/2}(E,C_{11/2}^{(sp)})$.    Hence, the corresponding
experimental astrophysical $S$-factors can be used as an independent
source of   reliable information about the ANC's for the $\alpha
+^3{\rm {He}}\rightarrow ^7{\rm {Be}}$(g.s.) and $\alpha +^3{\rm
{He}}\rightarrow^7{\rm {Be}}$(0.429 MeV).

The weighted means of the  ANC-values recommended by us for $^3{\rm
{He}}+\alpha \rightarrow^7{\rm {Be}}$(g.s.) and $^3{\rm
{He}}+\alpha\rightarrow ^7{\rm {Be}}$(0.429 MeV), obtained from all
of the experimental data presented in Figs.\ref{fig5}\emph{d} and
\ref{fig6}\emph{d}, are equal to $(C_{1\,\,3/2}^{exp})^2$=23.19$\pm$
1.37 fm$^{-1}$ and $(C_{1\,\,1/2}^{exp})^2$=15.73$\pm$ 1.02
fm$^{-1}$ ($C_{1\,\,3/2}^{exp}$=4.82$\pm$ 0.14 fm$^{-1/2}$ and
$C_{1\,\,1/2}^{exp}$=3.97$\pm$ 0.13 fm$^{-1/2}$).    One should note
that the values of $C_{1\,\,3/2}$ and $C_{1\,\,1/2}$ should not be
equal, in contrast with the assumption made in Ref.\cite{PD}. The
corresponding values of the NVC's are $\mid G_{1\,\,3/2}\mid^2$=
1.11$\pm$ 0.07 fm and $\mid G_{1\,\,1/2}\mid^2$= 0.75$\pm$ 0.05 fm.
As noted earlier   in paper \cite{Igam97}, the values ANC's (NVC's)
$C_{1\,\,3/2} ^2$=18.19 fm$^{-1}$ and $C_{1\,\,1/2}^2$=15.02
fm$^{-1}$ ( $\mid G_{1\,\,3/2}\mid^2$= 0.86 fm and $\mid
G_{1\,\,1/2}\mid^2$= 0.71 fm) were obtained from   the experimental
data analysis \cite{Osb82}. But we mentioned  above, in paper
\cite{Igam97}, firstly, the contribution of the nuclear interior
($r<$ 4.0 fm) to the calculated astrophysical $S$-factors was not
included and, secondly, the values of ANC's were obtained from the
analysis of the experimental data \cite{Osb82}, which has
considerable spread. It is seen that taking into account the
contribution of the nuclear interior and use the  experimental data
more accurate than those in Ref.\cite{Igam97} one can strongly
influence the extracted values of the ANC's for $^3{\rm
{He}}+\alpha\rightarrow ^7{\rm {Be}}$(g.s.) and $^3{\rm {He}}+\alpha
\rightarrow\,\,\,^7{\rm {Be}}$(0.429 MeV). A comparison of the
present result    and that obtained in paper \cite{Igam97} shows
that  the underestimation of the contribution both of the nuclear
interior and of the nuclear exterior indeed occurs in \cite{Igam97}
since the present value of ANC $C_{1\,\,3/2} ^2$ obtained from the
analysis of the more accurate experimental astrophysical $S$-factor
[6-10] is larger than that obtained in \cite{Igam97}.

The resulting ANC (NVC) values obtained by us are in   good
agreement with the values $C_{1\,\,3/2}^2$=20.52 fm$^{-1}$ and
$C_{1\,\,1/2}^2$=15.23 fm$^{-1}$ ($|G_{1\,\,3/2}|^2$= 0.97 fm and
$|G_{1\,\,1/2}|^2$= 0.72 fm) \cite{Wall84,Igam97}. However, the
results recommended by us for these ANC's differ noticeably from the
values $C_{1\,\,3/2}^2$=12.60$\pm$ 1.07 fm$^{-1}$  and
$C_{1\,\,1/2}^2=8.41\pm 0.58$ fm$^{-1}$($C_{1\,\,3/2}$=3.55$\pm$
0.15 fm$^{-1/2}$, $C_{1\,\,1/2}$=2.90$\pm$ 0.10 fm$^{-1/2}$,
$|G_{1\,\,3/2}|^2$=0.60$\pm$ 0.05 fm and
$|G_{1\,\,1/2}|^2$=0.40$\pm$ 0.03 fm) \cite{Nol01} as well as those
$C_{1\,\,3/2}^2=C_{1\,\,1/2}^2$=12.36 fm$^{-1}$
($C_{1\,\,3/2}=C_{1\,\,1/2}$=3.79 fm$^{-1/2}$ and
$|G_{1\,\,3/2}|^2=|G_{1\,\,1/2}|^2$=0.68  fm) \cite{PD}. In this
connection one would like to draw attention to the following. The
bound state wave functions and the initial state wave functions in
\cite{Nol01} were computed with different potentials and, so, these
calculations were not self-consistent. Besides, the values of the
binding energies for the bound states of $^7{\rm {Be}}$ calculated
in Ref.\cite{Nol01}    differ from the experimental ones. Therefore,
the calculated value of the binding energy  for the bound state of
$^7{\rm {Be}}$(g.s.) in the ($\alpha +^3{\rm {He}}$)-channel (4.73
MeV, see Table I in Ref.\cite{Nol01}) is also not in agreement with
the experimental one (1.59 MeV) . Since the ANC's (or NVC's) for  $
^3{\rm {He}}+\alpha \rightarrow ^7{\rm {Be}}$ are sensitive to the
form of the NN potential \cite{Kaj86}, it is desirable, firstly, to
calculate the wave functions of the bound state using other forms of
the NN potential, and, secondly, in order to  guarantee the
self-consistency, the same forms of the NN potential should be used
for such calculation of the initial wave functions. Besides, one
would also like to note  the recent result of Ref.\cite{PD} obtained
for $C_{1\,\,3/2}$ and $C_{1\,\,1/2}$ from the analysis of the
experimental $^3{\rm {He}}(\alpha,\gamma)^7{\rm {Be}}$ astrophysical
$S$-factors performed within the R-matrix method, where the
contribution from the internal part of the amplitude was simulated
by the background for a single pole. But there to reduce the number
of free parameters the assumption about equality of the ANC's
($C_{1\,\,3/2}=C_{1\,\,1/2}$) was used, and the best fitting of the
data was reached at $C_{1\,\,3/2}$=3.79 fm$^{-1/2}$ and the channel
radius $r_c$=3.0 fm.  It follows from here that in reality the
values of the ANC's, $C_{1\,\,3/2}$ and $C_{1\,\,1/2}$, should not
be equal. Moreover, the calculation shows that the asymptotic
behavior of the bound ($\alpha +^3{\rm {He}}$) state and $^3{\rm
{He}}\alpha$-scattering wave functions is reached, as it was
mentioned above, simultaneously only at $r_c\gtrsim$ 5.0 fm and, so,
at $r_c\ge$ 3.0 fm their substitution  for these wave functions in
the external part of the amplitude in Ref.\cite{PD} is not correct.

\subsection{$\alpha$-particle spectroscopic factors for the mirror
($^7{\rm {Li}}^7{\rm {Be}}$)-pair}

The $^{\glqq}$indirectly measured\grqq\, values of the ANC's for
$^3{\rm {He}}+\alpha\rightarrow ^7{\rm {Be}}$ obtained in the
present work and those for     $ \alpha +t\rightarrow ^7Li$ deduced
in Ref.\cite{Igam07} can be used  for obtaining information on the
ratio $R_{Z;j_f}=Z_{1j_f}(^7{\rm {Be}})/Z_{1j_f}(^7Li)$ for the
virtual $\alpha$ decays of the bound mirror ($^7Li^7{\rm {Be}}$)
-pair, where $Z_{1j_f}(^7{\rm {Be}}) (Z_{1j_f}(^7Li))$ is the
spectroscopic factor for $^7{\rm {Be}}$ ($^7Li$) in the ($\alpha
+^3{\rm {He}}$)(($\alpha +t$))-configuration. For this aim we can
easily derive the following from Eq.(\ref{13a})
\begin{equation}
R_{Z;\,j_f}=\frac{R_{C;\,j_f}}{R_{C^{(sp)};\,j_f}},
 \label{18acd}
\end{equation}
where $R_{C;\,j_f}=\Big (C _{1\,j_f}(^7{\rm
{Be}})/C_{1\,j_f}(^7Li)\Big)^2$($R_{C^{(sp)};\,j_f}=\Big (C
_{1\,j_f}^{(sp)}(^7{\rm {Be}})/C_{1\,j_f}^{(sp)}(^7Li)\Big)^2$) is
the ratio of squares of the ANC's (single-particle ANC's) for the
bound mirror ($^7Li^7{\rm {Be}}$)-pair and  $j_f$=3/2(1/2) for the
ground (first excited) state of the mirror nuclei. It should be
noted that in Eq.(\ref{18acd}) by using the values of the ANC's for
the $t+\alpha\to^7Li$ and $^3{\rm {He}}+\alpha\to^7{\rm {Be}}$
obtained in Ref.\cite{Igam07}  and in the present work,
respectively, one can verify a validity  of the approximation
($R_{C;j_f}\approx R_{C^{(sp)};\,j_f}$, i.e., $R_{Z;\,j_f}\approx$
1) used in Refs.\cite{Tim02,Tim07} for the mirror ($^7Li^7{\rm
{Be}}$) conjugated $\alpha$ decays. For the bound (first excited)
state of the mirror ($^7Li^7{\rm {Be}}$)-pair the ratio
$R_{C^{(sp)};\,3/2}$ ($R_{C^{(sp)};\,1/2}$) changes by only about
1.5\%(6\%)   under the variation of  the geometric parameters ($r_o$
and $a$) of the adopted Woods--Saxon potential \cite{Neu75, Neu83}
within the aforesaid ranges. The ratios are equal to
$R_{C^{(sp)};\,3/2} $=1.37$\pm$ 0.02 and
$R_{C^{(sp)};\,1/2}$=1.40$\pm$ 0.09. These ratios can be determined
by using the values of the single-particle ANC's for the mirror
($^7Li^7{\rm {Be}}$)-pair obtained in the present work and deduced
in Ref.\cite{Igam07}. The ratios for the ANC's are $ R_{C;\,3/2}
$=1.85$\pm$ 0.11 and $R_{C;\,1/2}$=1.75$\pm$ 0.10. From
(\ref{18acd}) the values of the ratio $R_{Z;\,j_f}$ are equal to
$R_{Z;\,3/2}$=1.35$\pm$ 0.08 and $R_{Z;\,1/2}$=1.25$\pm$ 0.11 for
the ground and the first excited states, respectively. These values
differ noticeably from those of $R_{Z;\,3/2}$=0.995$\pm$0.005 and
$R_{Z;\,1/2}$=0.99 calculated in Ref.\cite{Tim07} within the
microscopic cluster model. One of the reasons of these differences
can be associated with the  aforesaid approximation used in
Ref.\cite{Tim07}, which is hardly valid for the mirror ($^7Li^7{\rm
{Be}}$)-pair \cite{Tim05}.

Thus,  as  it is seen from here   that,  in fact, the magnitudes of
$R_{Z;\,j_f}$ differ noticeably from unity both for the ground state
and for the first excited state of the mirror ($^7Li^7{\rm {Be}}$)
-pair.

\subsection{Astrophysical $S$-factor for the
$^3{\rm {He}}(\alpha,\gamma )^7{\rm {Be}}$ reaction at solar
energies}

\hspace{0.7cm}  The equation (\ref{18}) and the weighted means of
the ANC's obtained  for the $^3{\rm {He}}+\alpha\to^7{\rm
{Be}}$(g.s) and $^3{\rm {He}}+\alpha\to^7{\rm {Be}}$(0.429 MeV)
can be used for calculating the $^3{\rm {He}}(\alpha,\gamma
)^7{\rm {Be}}$ astrophysical $S$-factor for capture to the ground
and first excited states as well as the total astrophysical
$S$-factor at solar energies ($E\leq 25$ keV). At first, we tested
again the fulfilment of the condition (\ref{15}) in the same way
as it is done above for $E\ge$ 90 keV. Similar results plotted in
Fig.\ref{fig1} are also observed for  the ${\cal
{R}}_{l_fj_f}(E,C^{(sp)}_{l_fj_f})$ function dependence    on the
single-particle ANC, $C^{(sp)}_{l_fj_f}$, at energies of $E< 90$
keV.

\begin{figure}[!h]
\epsfxsize=13.cm \centerline{\epsfbox{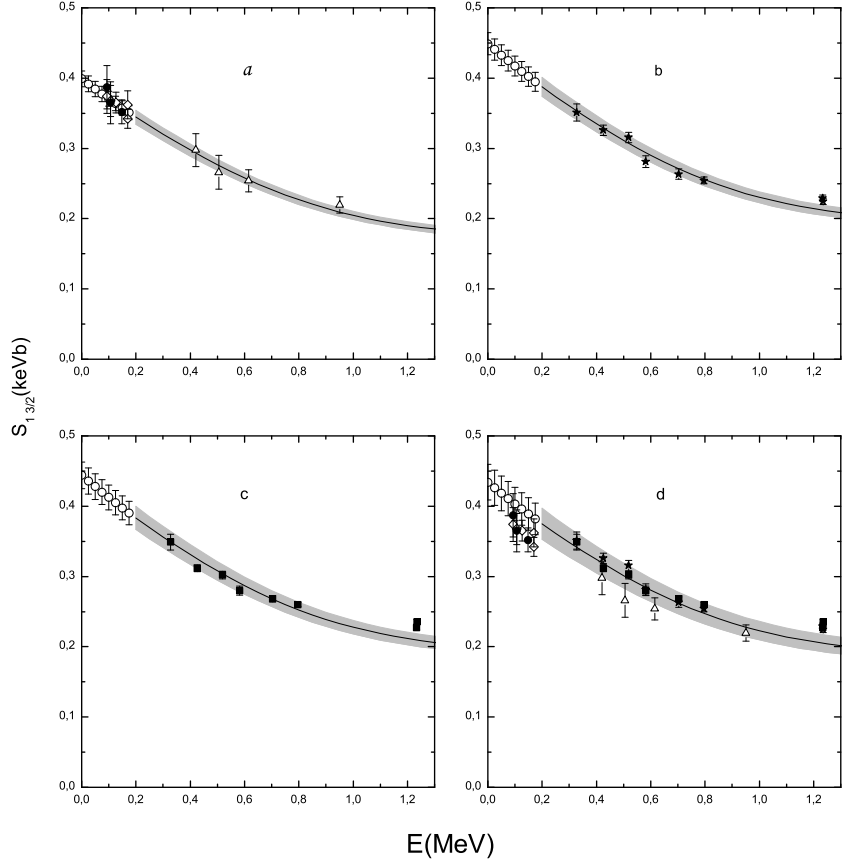}}
\caption{\label{fig7}The astrophysical $S$ - factors for the
$^3{\rm {He}}(\alpha,\gamma)^7{\rm
{Be}}$(g.s.)($(l_f,j_f)$=(1,3/2)) reaction. The opened diamond and
triangle symbols ($a$) are our result separated from  the total
experimental astrophysical $S$-factors of Refs.[7-9] and
\cite{Nara04}, respectively.  The filled circle symbols (filled
star and square symbols) are   experimental data of
Ref.\cite{Con07} (Ref.\cite{Brown07}, the activation ($b$) and the
prompt ($c$)). The opened   circle symbols are our results of the
extrapolation. The symbols in
 $d$ are data of all experiments \cite{Con07,Brown07} and the present work.
  The solid lines present our
calculations performed with the standard values of geometric
parameters $r_o$=1.80 fm and $a$=0.70 fm both for the bound
($\alpha +^3{\rm {He}}$) state and for $^3{\rm
{He}}\alpha$-scattering state. Everywhere the width of each of the
band is the weighted uncertainty.}
\end{figure}
\begin{figure}[!h]
\epsfxsize=13.cm \centerline{\epsfbox{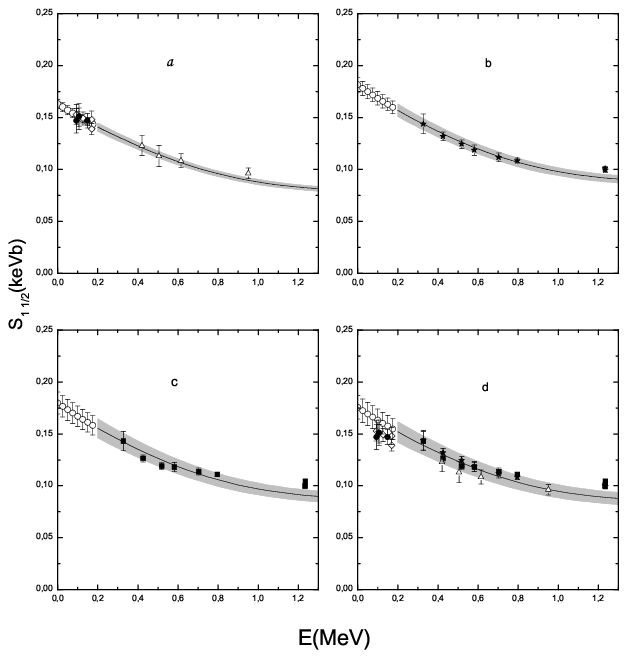}}
\caption{\label{fig8} The same as Fig.\ref{fig7} for the $^3{\rm
{He}}(\alpha,\gamma)^7{\rm {Be}}$(0.429
MeV)($(l_f,j_f)$=(1,1/2))reaction.}
\end{figure}

\begin{figure}[!h]
\epsfxsize=13.cm \centerline{\epsfbox{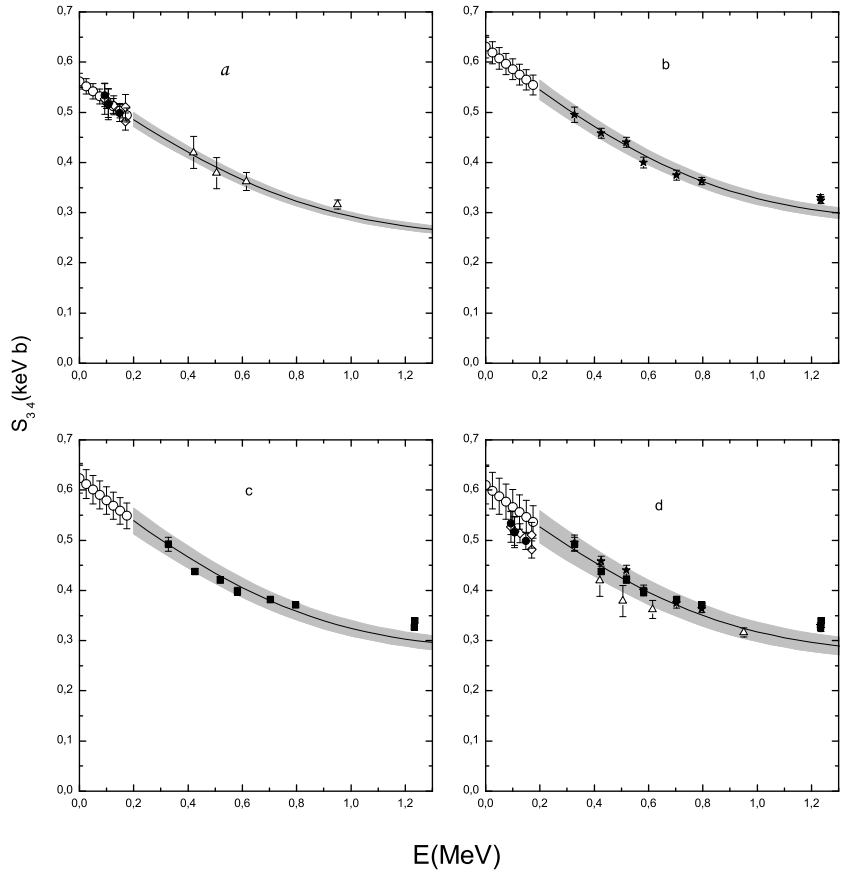}}
\caption{\label{fig9} The same as Fig.\ref{fig7} for the $^3{\rm
{He}}(\alpha,\gamma)^7{\rm {Be}}$(g.s.+0.429 MeV) reaction.}
\end{figure}

\begin{figure}[!h]
\epsfxsize=13.cm \centerline{\epsfbox{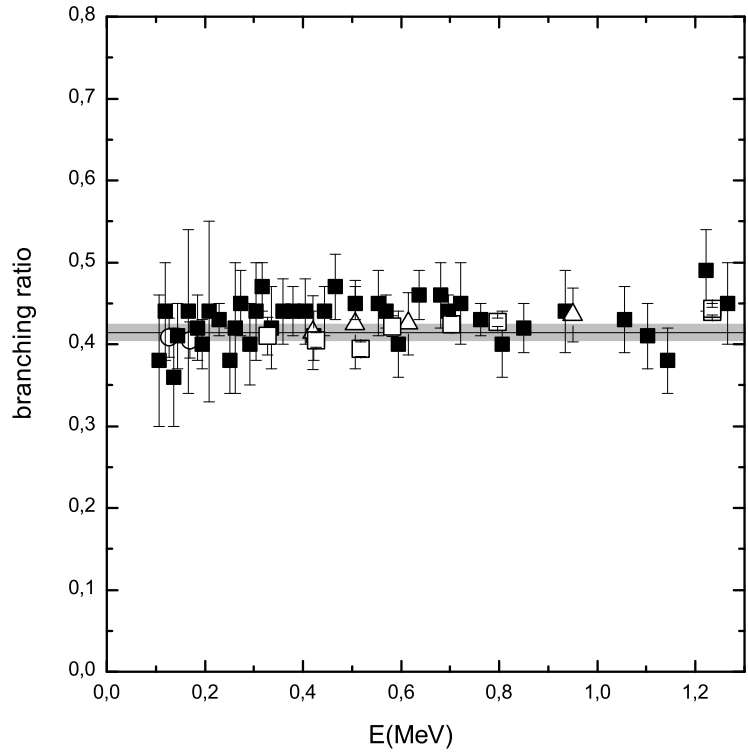}}
\caption{\label{fig10} The branching ratio. The filled square,
opened circle and    square  symbols are experimental data taken
from Refs.\cite{Kra82}, \cite{Con07} and  \cite{Brown07},
respectively, and the  opened triangle  symbols are our results.
The straight line and width of band are our results for the
weighted mean and its uncertainty, respectively.}
\end{figure}

The astrophysical $S$-factors for the    $^3{\rm
{He}}(\alpha,\gamma)^7{\rm {Be}}$(g.s.) and $^3{\rm
{He}}(\alpha,\gamma)^7{\rm {Be}}$(0.429 MeV) reactions are displayed
in Figs.\ref{fig7} and \ref{fig8}, respectively, as well as
presented in Table \ref{table1}. In Figs.\ref{fig7}$a$ and
\ref{fig8}$a$, the opened diamond and triangle symbols show our
result obtained from the analysis of the total experimental
astrophysical $S$-factors of [7-9] and \cite{Nara04}  by using the
corresponding values of the ANC's for each energy $E$ experimental
point presented in Table \ref{table1}. In Figs.\ref{fig7} and
\ref{fig8} ($a$,$b$ and $c$) the experimental data plotted by the
filled circle symbols (filled star and square symbols) are taken
from \cite{Con07} (from \cite{Brown07}). The symbols in
Figs.\ref{fig7}$d$ and \ref{fig8}$d$ are data of all experiments
\cite{Con07,Brown07} and our ones. The opened circle symbols in
these figures are the results of the extrapolation recommended in
the present work. The solid lines present our calculations performed
with the standard values of geometric parameters $r_o$=1.80 fm and
$a$=0.70 fm both for the bound ($\alpha +^3{\rm {He}}$) state and
for $^3{\rm {He}}\alpha$-scattering state. Everywhere the width of
each of the band is the weighted uncertainty, which includes the
uncertainties in the $^{\glqq}$indirectly measured\grqq\, values  of
ANC's    and the aforesaid uncertainty in the ${\cal
{R}}_{l_fj_f}(E,C^{(sp)}_{l_fj_f})$.
\newpage
\begin{landscape}
\begin{table}
\caption{\label{table1}The $^{\glqq}$indirectly measured\grqq\,
values of the asymptotic normalization constants (
$(C_{13/2}^{exp})^2$ and $(C_{11/2}^{exp})^2$) for $^3{\rm
{He}}+\alpha\to^7{\rm {Be}}$, the experimental astrophysical
$S$-factors ($S^{exp}_{1j_f}$ and $S^{exp}_{34}(E)$) and branching
ratio ($R^{exp}(E)$) at different energies E.}
 \begin{tabular}{|c|c|c|c|c|c|c|}\hline
$E$&\multicolumn{2}{c|}{$(C_{1\,\,j_f}^{exp})^2$}&\multicolumn{2}{c|}{$S^{exp}_{1\,\,j_f}$}&$S^{exp}_{34}(E)$&$R^{exp}(E)$ \\
 (keV)&\multicolumn{2}{c|}{(fm$^{-1})$}&\multicolumn{2}{c|}{(${\rm  {keV\,\,\,b}}$)}&(${\rm  {keV\,\,\,b}}$)& \\
 \cline{2-5}
&$j_f$=3/2&$j_f$=1/2&$j_f$=3/2&$j_f$=1/2&&\\ \hline
92.9&22.02$\pm$1.84&14.04$\pm$1.16&0.387$\pm$0.03\cite{Con07}&0.147$\pm$0.012\cite{Con07}&0.534$\pm$0.023\cite{Con07}&0.380$\pm$0.03\cite{Con07}\\\hline
93.3&21.41$\pm$1.38&14.66$\pm$0.94&0.374$\pm$0.02&0.153$\pm$0.009&0.527$\pm$0.03\cite{Con07}&0.409$\pm$0.03\\\hline
105.6&20.95$\pm$1.79&14.55$\pm$1.21&0.365$\pm$0.03\cite{Con07}&0.151$\pm$0.012\cite{Con07}&0.516$\pm$0.03\cite{Con07}&0.415$\pm$0.03\cite{Con07}\\\hline
106.1&21.23$\pm$1.28&14.55$\pm$0.88&0.368$\pm$0.02&0.150$\pm$0.009&0.518$\pm$0.03\cite{Con07}&0.408$\pm$0.02\\\hline
126.5&21.23$\pm$0.87&14.58$\pm$0.59&0.365$\pm$0.01&0.149$\pm$0.006&0.514$\pm$0.02\cite{Bem06}&0.408$\pm$0.02\\\hline
147.7&20.84$\pm$1.13&14.60$\pm$0.74&0.352$\pm$0.02\cite{Con07}&0.147$\pm$0.007\cite{Con07}&0.499$\pm$0.02\cite{Con07}&0.417$\pm$0.02\cite{Con07}\\\hline
168.9&20.57$\pm$0.81&14.09$\pm$0.55&0.343$\pm$0.01&0.139$\pm$0.006&0.482$\pm$0.02\cite{Bem06}&0.405$\pm$0.02\\\hline
170.1&21.90$\pm$1.21&15.00$\pm$0.83&0.362$\pm$0.02&0.148$\pm$0.008&0.510$\pm$0.02\cite{Con07}&0.409$\pm$0.03\\\hline
420.0&21.41$\pm$1.67&14.66$\pm$1.14&0.297$\pm$0.02&0.123$\pm$0.009&0.420$\pm$0.03\cite{Nara04}&0.414$\pm$0.05\\\hline
506.0&20.91$\pm$1.88&14.32$\pm$1.29&0.266$\pm$0.02&0.113$\pm$0.010&0.379$\pm$0.03\cite{Nara04}&0.424$\pm$0.05\\\hline
615.0&21.50$\pm$1.35&14.73$\pm$0.92&0.254$\pm$0.02&0.108$\pm$0.006&0.362$\pm$0.02\cite{Nara04}&0.425$\pm$0.04\\\hline
951.0&22.74$\pm$1.21&15.58$\pm$0.83&0.220$\pm$0.01&0.096$\pm$0.005&0.316$\pm$0.01\cite{Nara04}&0.436$\pm$0.03\\\hline
\end{tabular}
\end{table}
\end{landscape}
The results for   the total astrophysical $S$-factor $S_{34}(E)$ and
the branching ratio  for the reaction under consideration are
presented by Figs.\ref{fig9} and \ref{fig10}, respectively. The
opened circle symbols in Figs.\ref{fig9} are our result of
extrapolation in which each of the quoted uncertainties   is the
a.s.e., which involves the uncertainties both for the ANC's adopted
and that in ${\cal {R}}_{l_fj_f}(E,C^{(sp)}_{l_fj_f})$. The solid
lines and the width of each of the band are the same as in
Figs.\ref{fig7} and \ref{fig8}. As it is seen from Fig.\ref{fig9},
the equation (\ref{18}) allows us to perform a correct extrapolation
of the corresponding astrophysical $S$-factors at solar energies
when the corresponding ANC-values are known. In particular, the
values of
 the total astrophysical $S$-factor $S_{34}(E)$
 at solar energies are presented in Table \ref{table2}, and those  recommended
  by us are $S_{34}(0)=0.610 \pm 0.037$ ${\rm
{keV\,\, b}}$
 and $S_{34}$(23 keV)=0.599$\pm$ 0.036 ${\rm
{keV\,\, b}}$\footnote{The energy of $E$=23 ${\rm keV}$ corresponds
to the Gamow one.}.
\newpage
\begin{landscape}
\begin{table}
\caption{\label{table2}The weighted means of the ANC-values
($C^{{\rm exp}})^2$ for   $^3He +\alpha\rightarrow ^7{\rm Be}$,
NVC's $\mid G\mid^2_{{\rm exp}}$ and the calculated values of $
S_{3\,4}(E)$ at energies $E$=0 and 23 keV }
\begin{tabular}{|l|c|c|c|c|c|c|}\hline
Exp.&$(C^{{\rm exp}}_{1\,3/2})^2$, fm$^{-1}$&$\mid
G_{1\,3/2}\mid^2_{{\rm exp}}$, fm&$(C^{{\rm exp}}_{1\,1/2})^2$,
fm$^{-1}$&$\mid G_{1\,1/2}\mid^2_{{\rm exp}}$, fm&$S_{3\,4}$(0),
keV b&$S_{3\,4}$({\rm {23\,\,keV}}), keV b
\\ \hline \cite{Brown07}(the activation)&23.98$\pm $0.82&
1.14$\pm$0.04&16.25$\pm$0.62&0.77$\pm$0.03&0.630$\pm$0.022&0.619$\pm$0.022
\\ \hline
\cite{Brown07}(the
prompt)&23.72$\pm$1.01&1.13$\pm$0.05&16.09$\pm$0.94&0.76$\pm$0.05&0.624$\pm$0.029&$0.612\pm$0.019
\\ \hline
\cite{Nara04,Bem06,Con07}&21.31$\pm$0.61&1.06$\pm$0.03&14.59$\pm$0.40&0.69$\pm$0.02&0.562$\pm$0.016&0.552$\pm$0.015
\\ \hline
\cite{Nara04,Bem06,Con07,Brown07}
&23.19$\pm$1.37&1.11$\pm$0.07&15.73$\pm$1.02&0.75$\pm$0.05&0.610$\pm$0.037&0.599$\pm$0.036
\\ \hline
\end{tabular}
\end{table}
\end{landscape}
Comparison of our results with those of the authors of
Refs.\cite{Igam97} and \cite{Nol01} shows that a noticeable
discrepancy between the present results and those of
Refs.\cite{Igam97,Nol01} occurs. This circumstance is apparently
connected  with the underestimated value of $C_{1\,\,3/2}^2$ and
$C_{1\,\,1/2}^2$ (or $|G_{1\,\,3/2}|^2$ and $|G_{1\,\,1/2}|^2$)
obtained in Ref.\cite{Igam97,Nol01} in respect to our result.
However, our result is also in a good agreement with that
recommended in Refs.[4,6--10] and differs slightly from that
recommended in Refs.\cite{PD,Nara04} ($S_{34}(0)$=0.51$\pm$0.04 keV
b \cite{PD} and 0.53 ${\rm {keV\,\, b}}$ \cite{Nara04}).

Besides, we observe that the value of $S_{34}(0)$= 0.56 ${\rm
{keV\,\, b}}$ \cite{Lan86}   obtained within the microscopical
($\alpha +^3{\rm {He}}$)-cluster approach    is also in    agreement
with our result about 1.4$\sigma$ level. It follows from here that
the mutual agreement between the results obtained in the present
work and in \cite{Lan86}, which is based on the common approximation
about the cluster $(\alpha+^3{\rm {He}})$ structure of the $^7{\rm
{Be}}$, allows one to draw a conclusion  about  the dominant
contribution of the $(\alpha+^3{\rm {He}})$ clusterization to the
low-energy $^3{\rm {He}}(\alpha,\gamma)^7{\rm {Be}}$ cross section
both in the absolute normalization  and in the energy dependence
[6--10].  Therefore, single-channel $(\alpha+^3{\rm {He}})$
approximation for $^7{\rm {Be}}$ \cite{Lan86} is quite appropriate
for this reaction in the considered energy range.

One notes also that the ratios of  the $^{\glqq}$indirectly
measured\grqq\, astrophysical $S$-factors, $S_{1\,\,3/2}$(0) and
$S_{1\,\,1/2}(0)$, for the $^3{\rm {He}}(\alpha,\gamma )^7{\rm
{Be}}$ reaction populating to the ground and first excited states
obtained in the present work to those for the mirror
$t(\alpha,\gamma )^7{\rm {Li}}$ reaction populating to the ground
and first excited states deduced in Ref.\cite{Igam07} are equal to
$R_S^{(g.s.)}$=6.4$\pm$0.8 and $R_S^{(exc.s.)}$=6.1$\pm$0.7,
respectively.  These values are in a  good agreement with those of
$R_S^{(g.s.)}$=6.6 and $R_S^{(exc)}$=5.9 deduced in Ref.\cite{Tim07}
within the microscopic cluster model.

Fig.\ref{fig10}  shows a comparison  between the branching ratio
$R^{exp}(E)$ obtained in the present work (the opened   triangle
symbols) and that recommended in Refs.\cite{Kra82} (the filled
square symbols), in \cite{Con07} (the opened circles) and
\cite{Brown07} (the opened squares ). There the solid line and the
width of the band present the weighted mean $\bar{R}^{exp}$ of the
$R^{exp}(E)$ (the opened   triangle symbols) and the weighted
uncertainty obtained by us, respectively, which is equal to
$\bar{R}^{exp}$=0.41$\pm$ 0.01. As it is seen from
Fig.\ref{fig10}, the branching ratio obtained in the present work
and in \cite{Bem06,Brown07} is in a good agreement with that
recommended in Ref.\cite{Kra82} although the underestimation
occurs for the $S_{34}^{exp}(E)$ obtained in Ref.\cite{Kra82}.
Such a good agreement between two of the experimental data for the
$R^{exp}(E)$ can apparently be explained by the fact that  there
is a reduction factor  in \cite{Kra82}, being overall for the
$^3{\rm {He}}(\alpha,\gamma)^7{\rm {Be}}$(g.s.) and $^3{\rm
{He}}(\alpha,\gamma)^7{\rm {Be}}$(0.429 MeV) astrophysical
$S$-factors. The present result for $\bar{R}^{exp}$  is in a good
agreement with those of 043$\pm$ 0.02 \cite{Kra82} and 0.43
\cite{Moh93,Alt88} but  is noticeably larger than 0.37
\cite{Nol01} and 0.32$\pm$ 0.01 \cite{Sch87}.

Thus,  it follows  from here that the overall normalization of the
astrophysical $S$-factors at extremely low energies for the
reactions under consideration is mainly determined by the ANC -
values for the $^3{\rm {He}}+\alpha\rightarrow ^7{\rm {Be}}$(g.s.)
and $^3{\rm {He}}+\alpha\rightarrow ^7{\rm {Be}}$(0.429 MeV), which
can be determined rather well from the model independent analysis of
the precise experimental astrophysical $S$-factor [6--10], and the
values of the ANC's allow us to perform correct extrapolation of the
astrophysical $S$-factors for the direct radiative capture $^3{\rm
{He}}(\alpha,\gamma )^7{\rm {Be}}$ reaction at solar energies,
including $E$=0, and to predict the separated experimental
astrophysical $S$-factors for the $^3{\rm {He}}(\alpha,\gamma
)^7{\rm {Be}}$(g.s.) and $^3{\rm {He}}(\alpha,\gamma )^7{\rm
{Be}}$(0.429 MeV)  reactions at low experimentally acceptable energy
regions (126.5$\le$ E$\le$ 951 keV) obtained by using the total
experimental astrophysical $S$-factors measured in Refs.[6--9].

\section{Conclusion}

\hspace{0.7cm} The analysis of the experimental astrophysical
$S$-factors, $S^{exp}_{34}(E)$, for the $^3{\rm
{He}}(\alpha,\gamma)^7{\rm {Be}}$ reaction, which were precisely
measured at energies $E$=92.9-1235 keV [6--10], has been performed
within the modified two-body potential approach proposed recently in
Ref.\cite{Igam07}. The scrupulous quantitative analysis   shows that
the $^3{\rm {He}}(\alpha,\gamma)^7{\rm {Be}}$ reaction within the
considered energy ranges is peripheral and the parameterization of
the direct astrophysical $S$-factors in terms of ANC's for the
$^3{\rm {He}}+\alpha\rightarrow ^7{\rm {Be}}$ is adequate to the
physics of the peripheral reaction under consideration.

It is demonstrated that the experimental astrophysical $S$-factors
of the reaction  under consideration measured in the aforesaid
energy region    can be used as an independent source of getting
the information about the ANC's (or NVC's) for $^3{\rm
{He}}+\alpha\rightarrow ^7{\rm {Be}}$.  The weighted means of the
ANC's (NVC's) for   $^3{\rm {He}}+\alpha\rightarrow ^7{\rm {Be}}$
are obtained. They have to be $(C_{1\,\,3/2}^{exp})^2$=23.19$\pm$
1.37 fm$^{-1}$ and $(C_{1\,\,1/2}^{exp})^2$=15.73$\pm$ 1.02
fm$^{-1}$ for   $^3{\rm {He}}+\alpha\rightarrow ^7{\rm {Be}}$(g.s)
and   $^3{\rm {He}}+\alpha\rightarrow ^7{\rm {Be}}$(0.429 MeV),
respectively. The corresponding values of the NVC's are $\mid
G_{1\,\,3/2}\mid^2$= 1.11$\pm$ 0.07 fm and $\mid
G_{1\,\,1/2}\mid^2$= 0.75$\pm$ 0.05 fm. The uncertainty in the ANC
(NVC )-values includes the experimental errors for the
experimental astrophysical $S$-factors, $S^{exp}_{34}(E)$, and
that of the used approach. Besides, the values of   ANC's were
used for getting the information about the $\alpha$-particle
spectroscopic factors for the mirror ($^7Li^7{\rm {Be}}$)-pair.

The obtained values  of the ANC's  were also used for  obtaining
the experimental  $^3{\rm {He}}(\alpha,\gamma)^7{\rm {Be}}$
astrophysical $S$-factors for capture to the ground and first
excited states,  the branching ratio at the six experimental
points of energy $E$ ($E\ge$ 126.5 keV) [6--9] and for their
extrapolation at energies less than 90 keV, including $E$=0. In
particular, for the weighted mean of the branching ratio
$\bar{R}^{exp}$ and the total astrophysical $S$-factor $S_{34}(0)$
the values of $\bar{R}^{exp}$=0.41$\pm$ 0.01 and $S_{34}(0)=0.610
\pm 0.037$ keV b have been obtained, respectively. The latter is
noticeably larger than the result of $S_{34}(0)$=0.507$\pm$ 0.016
keV b , deduced in Ref.\cite{Adel98} from the measurements of
capture $\gamma$ ray, and is in an   agreement with those of
$S_{34}(0)$=0.572$\pm$ 0.026 keV b \cite{Adel98}, deduced  in
Ref.\cite{Adel98}  from the measurements of   $^7{\rm {Be}}$
activity, and $S_{34}(0)$=0.56 keV \cite{Lan86} obtained within
the microscopical ($\alpha +^3{\rm {He}}$)-cluster model.

\vspace{5mm} {\bf Acknowledgments} \vspace{5mm}

The authors  thank D.Bemmerer for providing the experimental results
of the updated data analysis .  The work has been supported by   The
Academy of Science  of The Republic of  Uzbekistan (Grant
No.FA-F2-F077).

\end{document}